\documentclass[letterpaper, 10 pt, conference]{ieeeconf}  

\IEEEoverridecommandlockouts                              

\usepackage{graphics} 
\usepackage{epsfig} 
\usepackage{amsmath} 
\usepackage{amssymb}  
\usepackage{multirow}
\usepackage{booktabs}
\usepackage{nameref}
\usepackage{soul}
\usepackage{xcolor}
\usepackage{array}
\usepackage{threeparttable}
\usepackage{cite}

\newcolumntype{P}[1]{>{\centering\arraybackslash}M{#1}}
\newcolumntype{M}[1]{>{\centering\arraybackslash}m{#1}}

\title{\LARGE \bf
Intermittent Control for Safe Long-Acting Insulin Intensification for Type 2 Diabetes: \textit{In-Silico} Experiments*
}

\author{Anas El Fathi$^{1}$, Mohammadreza Ganji$^{1}$, Dimitri Boiroux$^{2}$, Henrik Bengtsson$^{2}$, Marc D. Breton$^{1}$
\thanks{*This work was supported by UVA-Novo Nordisk Master Research Agreement.}
\thanks{$^{1}$Anas El Fathi, Mohammadreza Ganji, and Marc D. Breton are with the University of Virginia, Center for Diabetes Technology, Charlottesville, VA, USA
        {\tt\small \{fwt9vd, mex5yy, mb6nt\}@virginia.edu}}%
\thanks{$^{2}$Dimitri Boiroux and Henrik Bengtsson are with Novo Nordisk A/S, DK-2880, Bagsværd, Denmark
        {\tt\small \{dzbo, hbss\}@novonordisk.com}}%
}

\begin{document}

{\color{red} \textbf{© 2023 IEEE. Personal use of this material is permitted. Permission from IEEE must be obtained for all other uses in any current or future media, including
reprinting/republishing this material for advertising or promotional purposes, creating new
collective works for resale or redistribution to servers or lists, or reuse of any copyrighted
component of this work in other works.}}

\maketitle
\thispagestyle{empty}
\pagestyle{empty}

\begin{abstract}

Around a third of type 2 diabetes patients (T2D) are escalated to basal insulin injections. Basal insulin dose is titrated to achieve a tight glycemic target without undue hypoglycemic risk. In the standard of care (SoC), titration is based on intermittent fasting blood glucose (FBG) measurements. Lack of adherence and the day-to-day variabilities in FBG measurements are limiting factors to the existing insulin titration procedure. We propose an adaptive receding horizon control strategy where a glucose-insulin fasting model is identified and used to predict the optimal basal insulin dose. This algorithm is evaluated in \textit{in-silico} experiments using the new UVA virtual lab (UVlab), and a set of T2D avatars matched to clinical data (NCT01336023). Compared to SoC, we show that this control strategy can achieve the same glucose targets faster (as soon as week 8) and safer (increased hypoglycemia protection and robustness to missing FBG measurements). Specifically, when insulin is titrated daily, a time-in-range (TIR, 70--180 mg/dL) of 71.4$\pm$20.0\% can be achieved at week 8 and maintained at week 52 (72.6$\pm$19.6\%) without an increased hypoglycemia risk as measured by time under 70 mg/dL (TBR, week 8: 1.3$\pm$1.9\% and week 52: 1.2$\pm$1.9\%), when compared to the SoC (TIR at week 8: 59.3$\pm$28.0\% and week:52 72.1$\pm$22.3\%, TBR at week 8: 0.5$\pm$1.3\% and week 52: 2.8$\pm$3.4\%). Such an approach can potentially reduce treatment inertia and prescription complexity, resulting in improved glycemic outcomes for T2D using basal insulin injections.

\end{abstract}

\section{INTRODUCTION}

Type 2 diabetes (T2D) is a progressive condition characterized by insulin deficiency, insulin resistance, and increased hepatic glucose. People with T2D suffer from abnormally elevated blood glucose levels, causing an increased risk of morbidity (cardiovascular, rhinopathy, neuropathy, nephropathy, and others) and premature mortality \cite{dc23-S002}. T2D is initially treated with changes to lifestyle, such as diet and physical activity, in addition to oral anti-hyperglycemic agents such as Metformin, GLP-1 agonist, DPP-4 inhibitors, SGLT2 inhibitors, or others. Around a third of people with T2D ($\sim$100 million people worldwide) are escalated into insulin therapy, starting with long-acting insulin injections, also called basal insulin \cite{dc23-S009}.

The basal insulin dose is titrated to achieve a tight glycemic target without undue hypoglycemic risk \cite{dc23-S006}. The standard-of-care clinical approach to an insulin titration procedure involves adjusting the dose by a few units using fasting blood glucose (FBG) measurement. FBG can be measured with a capillary fingerstick (self-monitoring blood glucose (SMBG)) at fasting conditions, usually before the breakfast meal \cite{arnolds2013common}. Recent long-acting insulin (e.g., Glargine-300 and Degludec) have a duration effect of more than 24 hours, causing the steady state only to be achieved after a couple of days. As a result, basal insulin is usually titrated no more often than every three days (twice or once weekly) \cite{dc23-S009}, although daily titration has been proven feasible \cite{yale2017titration}. 

In the standard-of-care (SoC), for a chosen FBG target (e.g., 72--90 mg/dL), the individual is required to track their FBG measurements daily (e.g., using pre-breakfast SMBG) and follow a threshold-based rule as shown in table \ref{tab:rule}. This procedure is still challenging for multiple patients with documented lack of adherence \cite{yavuz2015adherence}. As a result, healthcare professionals are reluctant to insulin initiation, the so-called therapeutic inertia. Variability in insulin sensitivity between people with T2D and the day-to-day variability in FBG measurements also limit existing insulin titration procedures \cite{umpierrez2018glycemic}. It was estimated that $\sim$60\% of people with T2D using insulin do not reach recommended treatment targets \cite{wong2012comparison}.

\begin{table}[ht]
    \centering
    \caption{Protocol-specified titration algorithm for insulin Degludec as used by \cite{gough2014efficacy} in clinical trial NCT01336023.}
    \begin{tabular}{l|c | c}
       \toprule
       \textbf{Measurement}  &  \textbf{Thresholds mg/dL} & \textbf{Dose} \\
       \midrule
       Any pre-breakfast SMBG & $<$ 72 & -2 \\
       Mean pre-breakfast SMBG & $\geq 72$ $\&$ $\leq90$ & 0\\
       Mean pre-breakfast SMBG & $>$ 90 & 2\\
       \bottomrule
    \end{tabular}
    \label{tab:rule}
\end{table}

From a control systems perspective, the insulin titration procedure can be seen as an intermittent control problem where the actuator (insulin dose) can only be altered occasionally, the observed parameter (FBG) can be measured more frequently, and the control objective is to achieve glycemic targets in terms of mean FBG, HbA1c (a biomarker correlated with the mean blood glucose level over three months), and hypoglycemia protection \cite{battelino2022continuous}. In this work, we leverage the naive relationship between fasting glucose and insulin, that is, increasing insulin should decrease fasting glucose, to propose a receding horizon control (RHC) strategy for basal insulin titration that is robust to missing observations, faster to achieve steady-state, and safer in terms of hypoglycemia protection compared to the SoC. 

Although research in insulin control algorithms, and mainly in model predictive control (MPC), for type 1 diabetes (T1D) has flourished in the last decade \cite{el2018artificial}, fewer publications have targeted the T2D population. Arad{\'o}ttir et al. showed the feasibility of an MPC approach for basal insulin titration \cite{aradottir2019model}. While the MPC-based solution provided superior performance than the SoC, this work was a proof-of-concept where the same physiological parameters were shared between the simulator and the MPC internal model. Krishnamoorthy et al. proposed a model-free approach employing recursive least square-based extremum seeking control \cite{krishnamoorthy2020model}. They showed that this strategy converges to an insulin dose that steers the blood glucose concentration to a desired target (a control-to-target strategy compared to the SoC control-to-range strategy). The authors noted that testing their method on a more detailed physiological model was warranted. Other proprietary automated device-supported titration algorithms have been successfully evaluated in clinical trials \cite{davies2019randomized, bergenstal2019automated, hermanns2023evaluation, tews2022smartphone}, showing a trend in moving from paper-based SoC rules towards digital solutions \cite{thomsen2022time}. Additionally, other strategies in people with T1D may be extended to T2D exist and are clinically validated \cite{nimri2022comparison, bisio2022impact}, or under ongoing clinical investigation \cite{el2022titration, el2020model}.

In this manuscript, we propose a new formulation of RHC that: (i) employs a simple linear model with two parameters estimated online so that this model-based approach can be used in a broad population; (ii) drives FBG to a glucose target in the form of threshold interval as done in SoC, so this method is intuitively translated to clinical practice; (iii) organically accounts for day-to-day variability in FBG measurements; (iv) is evaluated in a large simulation platform with a detailed physiological model and avatars that reproduced a clinical trial; (v) is robust to missing FBG measurements, is safer in terms of hypoglycemia protection and is faster compared to the SoC. 

\section{METHODS}

\subsection{A Receding Horizon Formulation}

In RHC, a finite horizon optimal control problem is solved to generate a control trajectory. The resulting control trajectory is applied to the system for a fraction of the horizon length, usually one step. This process is then repeated, resulting in a practical strategy for using a model through online optimization while handling physical and desired constraints.

In the following, $t$ will indicate the time now and/or the discrete index of the time now. Let $y_t$ be the predicted FBG at time $t$, $z_t$ be the measured FBG, and $u_t=u_{t-1} + \delta u_{t}$ is the basal insulin dose given at time $t$ derived from the previous dose given at time $t-1$. We formulate the RHC problem in (\ref{eq:rhc}) with $\delta u_{t}$ as the optimized variable.

\begin{equation} \label{eq:rhc}
\begin{aligned}
    \min_{\delta u_t}    \quad & \gamma q_p(y_{t:t+T}) + r_p(\delta u_t) \\
    \textrm{s.t.} \quad & y_\tau = h_p( u_{1:\tau-1}), \, \tau = t, \dots, t+T \\
                  \quad & u_\tau = u_{t-1} + \delta u_{t}, \, \tau = t, \dots, t+T \\  
                  \quad & |\delta u_{t}| \leq \max \left(\delta u_{\min}, \beta u_{t-1} \right)
\end{aligned}
\end{equation}

In the above, $\delta u_t$ represents the optimal change in the insulin dose at time $t$. $h_p(.)$ is a closed-form solution linking previous inputs (insulin doses) $u_{1:\tau-1}$ to the predicted FBG $y_\tau$. Notice that, unlike standard formulations: 
\begin{itemize}
    \item $\delta u_t$ is kept unchanged within the prediction horizon, which is equivalent to a control horizon of size 1.
    \item The state is implicit, and there is no state filtering using new measurements (e.g., a Kalman filter); instead, the feedback mechanism will be ensured by adapting the function $h_p(.)$ using previous measurements $Z_{1:t}$ as shown in section ``\nameref{sec:param}''.
\end{itemize}

$q_p$ is the performance cost, and $r_p$ is a regularization term. $\gamma$ is a gain-tuning variable balancing the ratio between speed to objective and oscillation around the objective (performance-robustness tradeoff). $\beta$ is the maximum fraction of insulin change from the previous dose and is usually determined by healthcare professionals depending on the frequency of insulin titration (e.g., $\beta$=15\%). $\delta u_{\text{min}}$ is the minimum insulin dose change as determined by SoC (e.g., $\delta u_{\text{min}} = 1U$). $T$ is the prediction horizon. 

The model function $h_p(.)$ is described in (\ref{eq:model}).

\begin{equation} \label{eq:model}
    h_p(u_{1:\tau}) = p_0 - p_1 I(u_{1:\tau})
\end{equation}

where $I(.)$ represents the insulin pharmacokinetics function (rate of absorption from the injection site to the plasma) described in  ``\nameref{app1}''. $p_0$ (mg/dL) is an individual-specific parameter representing the average FBG at zero insulin. $p_1$ (mg/dL per mU/L) is another individual-specific parameter representing FBG sensitivity to plasma insulin concentration resulting from exogenous insulin injections.

\subsection{A Standard-of-Care Driven Cost Function}

In order to ensure equivalency between this method and the SoC, the cost functions are built to solve a control-to-range problem where the range can be determined by a health care professional as shown in table \ref{tab:rule}. In here, the upper and lower thresholds are denoted $\text{FBG}_U$ and $\text{FBG}_L$, respectively.

The performance cost $q_p(.)$ is separated into a hypoglycemia mitigation term $q^{\text{hypo}}_p(.)$ and a hyperglycemia mitigation term $q^{\text{hyper}}_p(.)$ described in (\ref{eq:q}).

\begin{multline}
    \label{eq:q}
    q_p(y_{t:t+T}) = \xi  \underbrace{\dfrac{1}{T} \sum_{\tau=t}^{t+T} \min\left(\dfrac{y_\tau}{\text{FBG}_L}e^{-\alpha p_2} - 1, 0\right)^2}_{q^{\text{hypo}}_p(y_{t:t+T})} \\
    + \underbrace{\dfrac{1}{T} \max\left( \begin{matrix} \sum_{\tau=t}^{t+T} \max\left(\dfrac{y_\tau}{\text{FBG}_L}e^{-\alpha p_2} - 1, 0\right)^2 & , \\ \sum_{\tau=t}^{t+T} \max\left(\dfrac{y_\tau}{\text{FBG}_U}e^{\alpha p_2} - 1, 0\right)^2 & \,\end{matrix} \right)}_{q^{\text{hyper}}_p(y_{t:t+T})}
\end{multline}

where $\xi$ is a weight that balances hypoglycemia and hyperglycemia risks. $p_2$ is the estimated standard deviation of the difference between measurements and predictions $|Z_{1:t} - Y_{1:t}|$; this variability can be associated with behaviors, such as an SMBG after meal consumption instead of fasting conditions, or metabolic, such as variation in insulin sensitivity between days. $\alpha$ is a parameter chosen to punish FBG measurements that fall within the interval $z_\tau \in \left[\begin{matrix} y_\tau e^{-\alpha p_2} & y_\tau e^{\alpha p_2} \end{matrix}\right]$. Assuming a log-normally distributed error ($z_t-y_t$) and a 5\%-95\% confidence interval, we set $\alpha$ such as $\mathcal{P}(|Z_{1:t} - Y_{1:t}| < \alpha) = 0.9$, or $\alpha=1.65$. 

\begin{figure}[htp]
\begin{center}
        \includegraphics[width=8.6cm]{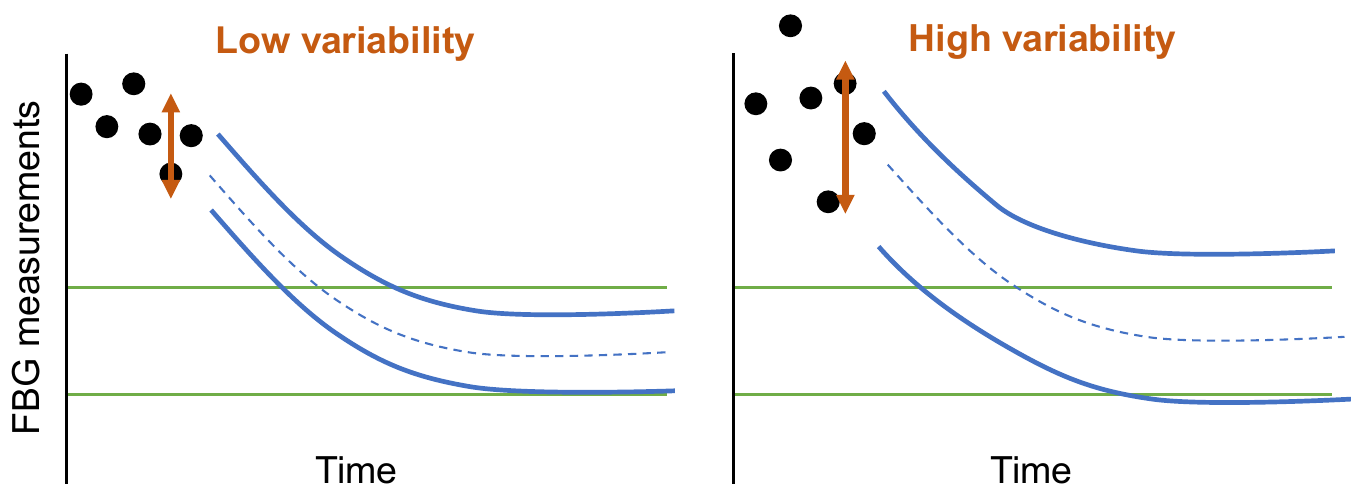}
\caption{Illustration of the expected trajectory resulted from optimizing the performance cost $q_p(.)$ in the cases of low variability (small $p_2$) and high variability (large $p_2$)} 
\label{fig:var_optim}
\end{center}
\end{figure}

$q^{\text{hypo}}(.)$ is a quadratic function that is non-zero only when $y_\tau e^{-\alpha p_2} < \text{FBG}_L$, meaning that the model is predicting future FBG measurements that might fall under $\text{FBG}_L$. Similarly, $q^{\text{hyper}}(.)$ is a quadratic function that is non-zero only when $y_\tau e^{-\alpha p_2} > \text{FBG}_L$ or $y_\tau e^{\alpha p_2} < \text{FBG}_U$, meaning that the model is predicting future FBG measurements that will not fall in the desired target $\text{FBG}_L$--$\text{FBG}_U$. Figure \ref{fig:var_optim} sketches an example of two trajectories resulting from optimizing the performance cost $q_p(.)$. 

The regularization term $r_p(.)$ punishes larger changes in the insulin dose as demonstrated in (\ref{eq:r})

\begin{equation}
\label{eq:r}
    r_p(\delta u_t) = \left( \dfrac{p_1}{p_{10}} \dfrac{\delta u_t}{\delta u_{\text{min}}} \right) ^2
\end{equation}

where $p_{1}$ is the estimated insulin sensitivity, while $p_{10}$ is the average insulin sensitivity. This formulation regularizes more sensitive individuals ($p_{1} > p_{10}$) and gives more freedom to change insulin doses for resistant individuals ($p_{1} < p_{10}$).

\subsection{Online parameter estimation} \label{sec:param}

Assuming that the parameter $p = \begin{bmatrix} p_0 & p_1 & p_2 \end{bmatrix}^\intercal$ comes from a known prior distribution, the maximum-a-posteriori estimate of $p$ (denoted as $p^*$) is the value for which $\mathcal{P}(p|Y_{1:t}, U_{1:t})$ attains its maximum, or

\begin{equation}
p^*=\text{arg}\,\max\limits_{p \in \mathbf{P}}\, \mathcal{P}(Y_{1:t} | U_{1:t}, p) \mathcal{P}(p)
\end{equation}

Given the assumption that errors in the fasting measurement $z_t - y_t$ follow a log-normal distribution with a zero mean and standard deviation of $p_2$, are independent and identically distributed, and the prior distribution of $p$ has a mean of $p_0 = \begin{bmatrix} p_{00} & p_{10} & p_{20} \end{bmatrix}^\intercal$, fixed standard deviation $\eta = \begin{bmatrix} \eta_{0} & \eta_{1} & \eta_{2} \end{bmatrix}^\intercal$ and no correlations, we can write:

\begin{multline} \label{eq:adapt}
p^*=\text{arg}\,\min\limits_{p \in \mathbf{P}}\, \dfrac{1}{p_2^2} \sum_{k=1}^{t} \log\left(\dfrac{z_k}{h_p(u_{1:k})}\right)^2  - 2 t \log \left(p_2\right) \\
+ \sum_{i=0}^{2} \dfrac{1}{\eta_i^2} \log\left(\dfrac{p_i}{p_{i0}}\right)^2
\end{multline}

Eq. \ref{eq:adapt} is resolved in each cycle when a new FBG is taken, and the parameter $p$ is used in the RHC formulation \ref{eq:rhc}.

\subsection{Experimental Setup}

The UVA virtual lab (UVlab) is a new simulator platform for T2D metabolic and behavioral replay equipped with a population of n=6156 \textit{in-silico} subjects (avatars) accounting for the heterogeneity and different subtypes of T2D. A subpopulation of n=427 avatars was extracted to match baseline characteristics and outcomes of the clinical trial NCT01336023 \cite{gough2014efficacy} (``\nameref{app2}'').

Five experiments were conducted: (i) SoC-3, where the standard of care in clinical practice is used: insulin is titrated twice weekly using the FBG for three consecutive days before dose titration, titration rules were the same as in table \ref{tab:rule}; (ii) SoC-1, which is the same as SoC-3, but only one FBG is used, the one at the day of dose titration; This scenario represents a highly non-adherent person where they only remember to take FBG when they need to change the insulin dose (iii) RHC-3, is similar to SoC-3, but the dose is computed using the proposed RHC control strategy; (iv) RHC-1, is similar to SoC-1, but the dose is computed using the proposed RHC control strategy; (v) RHC-1-acc, which is a novel approach where the insulin dose is titrated daily following the FBG measurement.

In all experiments, avatars consumed personalized meals consisting of three main meals and one snack with a ratio of 0.3/0.3/0.3/0.1 of their total daily carbohydrates. A fasting variability that mimics observed metabolic variability in clinical trials was introduced by daily perturbing the natural glucose steady state for each avatar. Variability was set to generate a 20 mg/dL standard deviation in fasting glucose. Avatars measured their FBG before the breakfast meal when required and injected insulin Degludec as their daily insulin basal. Real-time glucose was measured by a simulated continuous glucose monitoring (CGM) sensor. Simulations were executed for 52 weeks. 

In addition to clinical target parameters in table \ref{tab:rule}, the RHC control strategy uses population-level hyper-parameters described in table \ref{tab:hyper}.

\begin{table}[ht]
    \centering
    \caption{Hyper-parameter used in the experimental setup.}
    \begin{tabular}{c| l | c}
       \toprule
       \textbf{Symbol}  &  \textbf{Description} & \textbf{Value} \\
       \midrule
        $\gamma$ & Gain tradeoff performance-robustness & 250 \\
        $\xi$ & Gain tradeoff hypoglycemia-hyperglycemia & 100 \\
        $T$ & Prediction horizon in days & 10 \\
        $p_{00}$, $\eta_0$ & FBG at zero-insulin (mg/dL) & 150, 0.25\\
        $p_{01}$, $\eta_1$ & FBG sensitivity to insulin (mg/dL per mU/L) & 5.0, 0.5\\
        $p_{02}$, $\eta_2$ & FBG variability (unitless) & 0.15, 1.0\\
       \bottomrule
    \end{tabular}
    \label{tab:hyper}
\end{table}

\subsection{Outcomes metrics}
\label{sec:outcomes}
Algorithm performance was assessed using established metrics of quality of glycemic control in T2D \cite{battelino2022continuous, dc23-S006}, including CGM time $<70$ mg/dL (TBR); CGM time in the target $70-180$ mg/dL (TIR); HbA1c as computed by the Glycemic Management Index (GMI) formulation; mean FBG; Number of FBG $<54$ mg/dL indicating level-2 hypoglycemic risk; and total basal insulin (U). Metrics were calculated every 14 days.

\section{RESULTS}

\begin{figure}[htp]
\begin{center}
        \includegraphics[width=8.6cm]{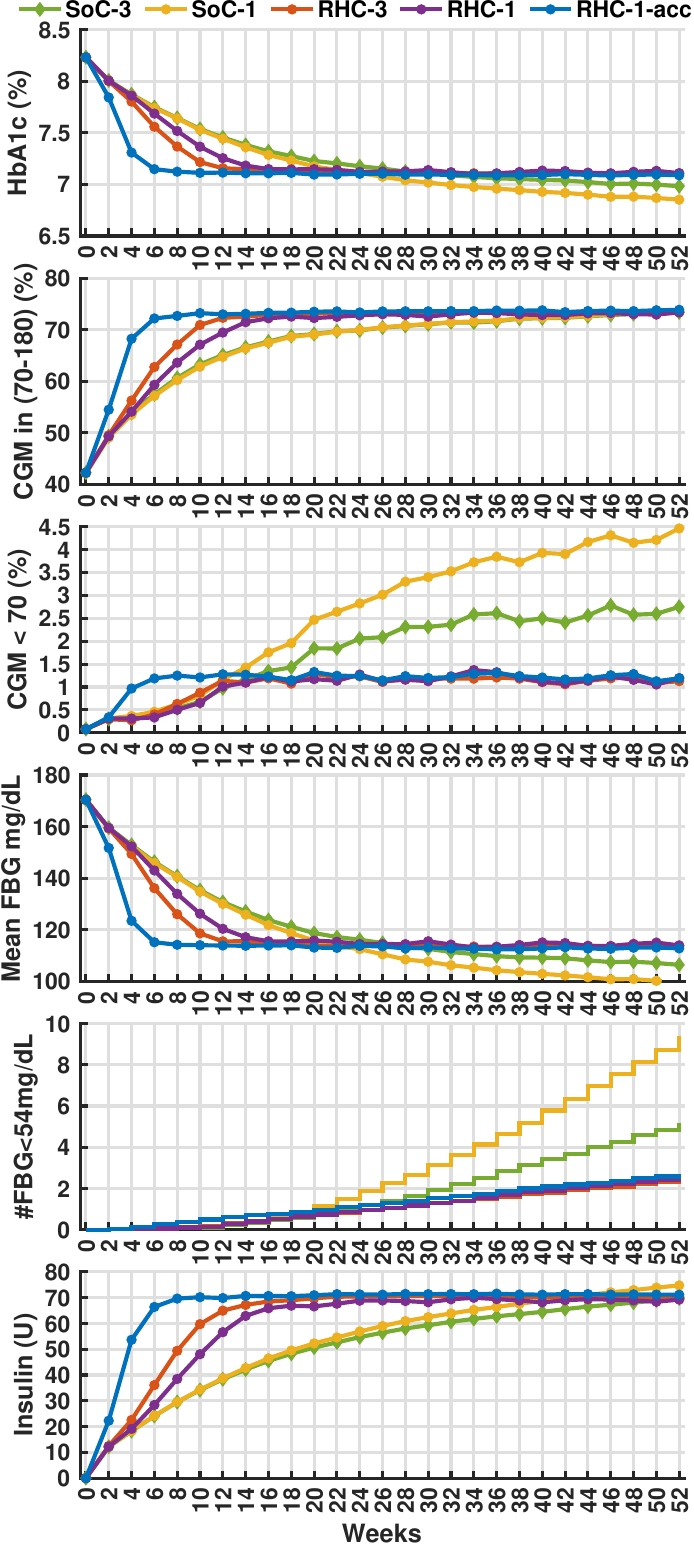}
\caption{Summary results of all the experiments: SoC-3 represent the standard clinical of care where insulin is titrated twice weekly using the FBG for three consecutive days prior to dose titration; SoC-1 is the same as SoC-3, but only one FBG is used, the one at the day of dose titration; RHC-3 uses the same protocol as SoC-3, but the dose is computed using the proposed RHC approach; RHC-1 is the same as RHC-3, but only one FBG is used, the one at the day of dose titration; RHC-1-acc is a proposed new approach where each day FBG is measured, and the dose is titrated. '\#' denotes 'number of.' Values are mean.} 
\label{fig:exp_all}
\end{center}
\end{figure}

\begin{table*}
\setlength\tabcolsep{3.1pt}
\begin{center}
  \begin{threeparttable}[b]
\caption{Percentage of avatars (n=427) achieving glycemic targets as defined by: Fasting target, the mean FBG is within 80mg/dL and 130mg/dL; HbA1C target, HbA1c $<$ 7\% with no level 2 hypoglycemic (FBG$<$54mg/dL); CGM target, TIR$>$70\% and TBR$<$4\%.}\label{tab:outcomes}
    \begin{tabular}{p{1.3cm}|M{0.86cm}M{0.86cm}M{0.86cm}M{0.86cm}M{0.86cm}|M{0.86cm}M{0.86cm}M{0.86cm}M{0.86cm}M{0.86cm}|M{0.86cm}M{0.86cm}M{0.86cm}M{0.86cm}M{0.86cm}|}
    \toprule 
             & \multicolumn{5}{c|}{Week 8} & \multicolumn{5}{c|}{Week 26} & \multicolumn{5}{c|}{Week 52} \\
             & \textbf{SoC-3} & \textbf{SoC-1} & \textbf{RHC-3} & \textbf{RHC-1} & \textbf{RHC-1\newline-acc} & \textbf{SoC-3} & \textbf{SoC-1} & \textbf{RHC-3} & \textbf{RHC-1} & \textbf{RHC-1\newline-acc} & \textbf{SoC-3} & \textbf{SoC-1} & \textbf{RHC-3} & \textbf{RHC-1} & \textbf{RHC-1\newline-acc} \\ 
\midrule
\textbf{Fasting\newline Target} & 49.1\newline($\pm$50.0) & 48.2\newline($\pm$50.0) & 70.4\newline($\pm$45.7) & 56.6\newline($\pm$49.6) & \emph{85.4\newline($\pm$35.3)} & 84.5\newline($\pm$36.2) & 82.3\newline($\pm$38.2) & 
\emph{87.6\newline($\pm$33.0)}  & 84.3\newline($\pm$36.4) & 87.3\newline($\pm$33.3) & \emph{91.8\newline($\pm$27.5)}  & 85.1\newline($\pm$35.6) & 88.3\newline($\pm$32.2) & 86.6\newline($\pm$34.1) & 89.2\newline($\pm$31.1)  \\ 
\textbf{HbA1c\newline Target} & 24.2\newline($\pm$42.9) & 24.6\newline($\pm$43.1) & 33.1\newline($\pm$47.1) & 26.3\newline($\pm$44.1) & \emph{40.6\newline($\pm$49.2)}	& 42.7\newline($\pm$49.5) & 38.3\newline($\pm$48.7) & 43.7\newline($\pm$49.7) & 43.2\newline($\pm$49.6) & 
\emph{45.1\newline($\pm$49.8)} & 38.5\newline($\pm$48.7) & 32.4\newline($\pm$46.9) & 43.2\newline($\pm$49.6) & 43.4\newline($\pm$49.6) & \emph{44.1\newline($\pm$49.7)}   \\ 
\textbf{CGM\newline Target} & 39.4\newline($\pm$48.9) & 39.7\newline($\pm$49.0) & 48.4\newline($\pm$50.0) & 43.0\newline($\pm$49.6) & \emph{51.2\newline($\pm$50.0)} & 48.6\newline($\pm$50.0) & 40.7\newline($\pm$49.2) & 52.8\newline($\pm$50.0) & 51.4\newline($\pm$50.0) & \emph{53.3\newline($\pm$50.0)}  & 45.5\newline($\pm$49.9) & 32.2\newline($\pm$46.8) & 52.6\newline($\pm$50.0) & 51.6\newline($\pm$50.0) & \emph{53.3\newline($\pm$50.0)}   \\ 
\bottomrule
    \end{tabular}
    \begin{tablenotes}
       \item Values are mean and standard deviations. Best values are highlighted in italics.
\end{tablenotes}
  \end{threeparttable}
\end{center}
\end{table*}

The results of the five experiments (SoC-3, SoC-1, RHC-3, RHC-1, and RHC-1-acc) are summarized in figure \ref{fig:exp_all}.

When using the SoC, averaging three consecutive days prior to dose titration is needed to minimize hypoglycemia exposure, mainly at week 52, SoC-3 \textit{vs.} SoC-1: CGM $<$ 70 mg/dL is 2.8$\pm$3.4 \textit{vs.} 4.5$\pm$4.6. This is not true for the proposed RHC approach where RHC-3 \textit{vs.} RHC-1: CGM $<$ 70 mg/dL (\%) is 1.1$\pm$1.9 \textit{vs.} 1.2$\pm$2.0. 

While a steady state in insulin dose does not seem to be achieved using SoC, with RHC, a steady state is achieved and is the same regardless of the number of FBG (three \textit{vs.} one) or the frequency of titration (twice-weekly \textit{vs.} daily). Nevertheless, the final insulin dose achieved at week 52 is similar across the five experiments; interestingly, there is a higher variation in insulin requirements in the RHC algorithms compared to SoC: SoC-3 \textit{vs.} SoC-1 \textit{vs.} RHC-3 \textit{vs.} RHC-1 \textit{vs.} RHC-3-acc: insulin dose (U) is 71.5$\pm$50.7 \textit{vs.} 75.8$\pm$50.3 \textit{vs.} 73.4$\pm$85.3 \textit{vs.} 72.7$\pm$81.3 \textit{vs.} 74.6$\pm$85.9). This is inverted for glycemic outcomes; less variability is seen with RHC compared to SoC, e.g., mean FBG (mg/dL) is 106.4$\pm$23.9 \textit{vs.} 99.4 $\pm$25.4 \textit{vs.} 113.7$\pm$13.7 \textit{vs.} 113.9$\pm$14.7 \textit{vs.} 112.8$\pm$13.6. This desired behavior indicates an individualization of the insulin dose to achieve the same target.

The RHC approach is generally faster in achieving its final insulin dose and titrating FBG regardless of the number of FBG and titration frequency. At week 8, the HbA1c (\%) is 7.6$\pm$1.0 \textit{vs.} 7.6$\pm$1.0 \textit{vs.} 7.4$\pm$0.7 \textit{vs.} 7.5$\pm$0.9 \textit{vs.} 7.1$\pm$0.6. This is corroborated in table \ref{tab:outcomes}, where the percentage of avatars achieving glycemic targets is presented. Across the targets, the RHC control strategy is faster at achieving targets and can converge as soon as week 8 when insulin is titrated daily ($\sim$90\% in target compared to $\sim$50\%).

\section{DISCUSSION}

The use of basal insulin is associated with improved glycemic outcomes in people with T2D. However, titrating basal insulin can be a prolonged and challenging process. This manuscript describes an adaptive algorithm based on RHC that has the potential to provide faster and safer insulin titration compared to the current SoC. This algorithm can be implemented as a smartphone application, making it accessible to patients. The smartphone application will require the same information as the SoC, including glycemic targets outlined in Table \ref{tab:rule} and measured FBG levels.

In the current SoC, day-to-day variability is compensated by taking three consecutive FBG measurements and averaging the results. However, this approach is susceptible to patient non-adherence, i.e., missing FBG measures. Furthermore, individuals with substantial day-to-day variability in FBG levels remain at risk for hypoglycemia, as even if the mean FBG is within the target range, there is no guarantee that a hypoglycemic event (low FBG) will not occur in the future. In the proposed formulation of the RHC problem, the day-to-day variability of FBG is estimated online using all historical FBG measures. Using the estimated variability envelope, the cost function forces an optimal trajectory that seeks the lowest possible FBG while maintaining a safe hypoglycemia profile (Figure \ref{fig:var_optim}). As a result, this strategy guides individuals with high and low variability to personalized glycemic targets.

Similarly, in the current SoC, the fact that changes in basal insulin dosing only results in a change in the steady state in 3 to 5 days is accommodated by only titrating after a minimum of a 3-day interval. The RHC formulation addresses this insulin pharmacokinetic delay by directly incorporating it into the model as expressed in (\ref{eq:model}). This simple model is critical to unlocking the tight and precise control of FBG following daily insulin titration, as presented in the RHC-1-acc experiment. Additionally, this simple formulation enables the extension of this approach to other basal insulin formulations (e.g., weekly insulin) by modifying the pharmacokinetics equation $I(u_{1:\tau})$. It is important to note that the \textit{in-silico} experiments employed a more comprehensive pharmacokinetics model and a more sophisticated insulin-glucose pharmacodynamics model, reflecting the ability of this simple model to predict FBG.

Another aspect of this problem is that although FBG measurements provide information about fasting glycemia for people with T2D, they do not entirely represent overall glycemia. This is reflected in table \ref{tab:outcomes} where fasting targets can be achieved by most of the population ($\sim$90\%) by week 52, yet around half do not achieve HbA1c or CGM targets. Compared to SoC, the RHC approach seems to increase this proportion by an additional 6\%-9\%. This can be explained by the personalization built into the RHC formulation, as people with low day-to-day variability in FBG can be brought to the lower edge of the target range. 

The presented approach relied on SMBG, which is still popular with people with T2D. However, CGM's recent affordability and availability may result in some patients using CGM instead of SMBG. Because of how it is formulated, the same algorithm can be used with any fasting measurement source. The CGM signal can be processed to extract fasting and used with this approach.

This work is limited by the proposed experimental setup, which can differ from real-world situations and was not examined in this manuscript. Notably,  physicians and/or people with T2D will likely stop titrating when a clear hypoglycemia trend is observed. As a result, the hypoglycemia risk of SoC-3 and Soc-1 beyond week 26 might be overestimated. Moreover, the model parameters were assumed constant over one year; in real-life, insulin sensitivity or day-to-day variability might change over time. 

\section{CONCLUSIONS}

This work examined the problem of insulin titration using intermittent FBG for people with T2D. A novel receding horizon control strategy is shown to be robust to missing FBG measurements and can converge to a steady state faster than the current SoC while ensuring a safe hypoglycemia profile. This algorithm can be implemented as a smartphone application and used by people with T2D. Such an approach can simplify the treatment and may improve adherence, resulting in improved glycemic outcomes. Results remain limited by the conducted \textit{in-silico} experiments, and real-world clinical studies using such an adaptive and  personalized control approach are warranted to confirm results. 





\section*{APPENDIX 1: Generic insulin absorption model} \label{app1}

Different formulations of basal insulin are available in the market. We express the pharmacokinetics of basal insulin in the following generic closed-form:

\begin{multline} 
    I(u_{1:\tau}) = 1000 \dfrac{F k_2  k_1}{k_{BW}  V_i  k_{cl}  (k_2-k_1 )} \\ \sum_{k=1}^{\tau} \left(e^{-k_1 (t_\tau-t_k) }-e^{-k_2 (t_\tau-t_k) }\right) u_k    
\end{multline}

where $F$, $V_i$, $k_{cl}$ $k_1$, $k_2$ are drug-specific parameters described in table \ref{tab:paramDegludec}. In this formulation, the parameter $k_1$ is associated with insulin half-time as $t_{1/2} = \dfrac{1}{k_1}$. Additionally, when basal insulin is injected with a fixed frequency $T_{\text{inj}}$, $k_2$ is related to the time-to-peak absorption as $t_{\text{max}} = \dfrac{\log \dfrac{k_1}{k_2} - \log \dfrac{1 - \exp\left(-k_1 T_{\text{inj}}\right)}{1 - \exp\left(-k_2 T_{\text{inj}}\right)}}{k_1 - k_2}$.

\begin{table}[ht]
\setlength\tabcolsep{3.8pt}
\begin{center}
\caption{\label{tab:paramDegludec} Generic insulin absorption model parameters for different insulin basal formulations.}
\begin{tabular}{lp{2.2cm}cccl}
\toprule
\textbf{Symbol} & \textbf{Description} & \textbf{Glar-100} & \textbf{Glar-300} & \textbf{Deg} & \textbf{Units} \\
\midrule
\textbf{$F$} & Bioavailability & 1.0 & 1.0 & 1.0 & unitless \\
\textbf{$V_i$} & Insulin distribution volume & 0.1 & 0.1 & 0.1 & L/kg \\
\textbf{$k_{cl}$} & Insulin clearance & 0.18 & 0.22 & 0.20 & 1/min \\
\textbf{$k_1$} & Time constant & 0.00067 & 0.00057 & 0.00068 & 1/min \\
\textbf{$k_2$} & Time constant & 0.0059	& 0.0019 & 0.0024 & 1/min \\
\bottomrule
\end{tabular}
\end{center}
\end{table}

\section*{APPENDIX 2: Matching of clinical data} \label{app2}
The UVlab has a set of avatars (n=6156) matched 1-to-1 to real people with T2D (unpublished data). For the proposed experimental setup, a subpopulation of n=427 avatars was selected to match the Degludec arm of the clinical trial NCT01336023 \cite{gough2014efficacy}. In addition to body weight, body mass index, age, and duration of diabetes, table \ref{tab:match} shows the matched glycemic metrics.

\begin{table}[ht]
  \begin{threeparttable}[b]
\setlength\tabcolsep{4.0pt}
\begin{center}
\caption{\label{tab:match} Metrics matched to real data.
}
\begin{tabular}{l|M{1.2cm}M{1.2cm}|M{1.2cm}M{1.2cm}}
\toprule
& \multicolumn{2}{c|}{\textbf{Baseline}} & \multicolumn{2}{c}{\textbf{Week 26}} \\
\textbf{Metrics} & \textbf{Data\newline(n=413)} & \textbf{Sim\newline(n=427)} & \textbf{Data\newline(n=413)} & \textbf{Sim\newline(n=427)} \\
\midrule
\textbf{HbA1c (\%)} & 8.3$\pm$1.0 & 8.3$\pm$1.0 & 6.9$\pm$1.1 & 6.9$\pm$1.1\\
\textbf{FBG (mg/dL)} & 169$\pm$49 & 169$\pm$49 & 105$\pm$41 & 105$\pm$41\\
\textbf{Insulin}     &  0$\pm$0      & 0$\pm$0       & 52 & 52$\pm$29 \\
\textbf{\#FBG $<$ 54mg/dL} & N/A & N/A & 1.2 & 1.2$\pm$1.8\\
\bottomrule
\end{tabular}
\begin{tablenotes}
       \item '\#' denotes 'number of.' Values are mean and standard deviation. 
\end{tablenotes}
\end{center}
\end{threeparttable}
\end{table}



\bibliographystyle{IEEEtran}
\bibliography{my}

\end{document}